\documentclass[aps,prb,twocolumn,showpacs]{revtex4}
\usepackage{graphicx}

\begin{document}

\title{Unexpected Weak Spatial Variation of Local Density of Sates Induced by Individual Co Impurity Atoms in Na(Fe$_{1-x}$Co$_x$)As Revealed by Scanning Tunnelling Spectroscopy}

\author{Huan Yang$^{1}$, Zhenyu Wang$^{2}$, Delong Fang$^1$, Sheng Li$^1$, Toshikaze Kariyado$^3$, Genfu Chen$^2$, Masao Ogata$^3$, Tanmoy Das$^4$ and A. V. Balatsky$^4$ and Hai-Hu Wen$^{1,*}$}

\affiliation{$^1$Center for Superconducting Physics and Materials, National Laboratory of Solid State Microstructures and Department of Physics, Nanjing University, Nanjing 210093, China}
\affiliation{$^2$National Laboratory for Superconductivity, Institute of Physics and National Laboratory for Condensed Matter Physics, Chinese Academy of Sciences, Beijing 100190, China}
\affiliation{$^3$Department of Physics, University of Tokyo, 7-3-1 Hongo, Bunkyo, Tokyo 113-0033 and JST, TRIP, Sanbancho, Chiyoda, Tokyo 102-0075, Japan}
\affiliation{$^4$Theoretical Division and Center for Integrated Nanotechnologies, Los Alamos National Laboratory, Los Alamos, NM, 87545, USA}

\begin{abstract}
We use spatially resolved scanning tunneling spectroscopy in
Na(Fe$_{1-x}$Co$_x$)As to investigate the impurity effect induced
by Co dopants. The Co impurities are successfully identified, and
the spatial distributions of local density of state at different
energies around these impurities are investigated. It is found
that the spectrum shows negligible spatial variation at different
positions near the Co impurity, although there is a continuum of
the in-gap states which lifts the zero-bias conductance to a finite
value. Our results put constraints on the S$\pm$ and S++ models
and sharpen the debate on the role of scattering potentials
induced by the Co dopants.
\end{abstract}

\maketitle
\section{Introduction}
According to the Anderson's theorem
\cite{Anderson,BalatskyRevModPhys}, the Cooper pairs with singlet
pairing can survive in the presence of non-magnetic impurities,
while the magnetic impurities are detrimental to superconductivity
yielding a unique pattern of the electronic local density of
states (LDOS). This view has been well established in the
conventional superconductors\cite{YazdaniScience}. In
superconductors with gap nodes, even the nonmagnetic impurities
can be strong pair
breakers\cite{BalatskyRevModPhys,PanZnImpurity}. In the iron-based
superconductors \cite{HosonoFeAs}, the pairing mechanism remains
as a hot and unresolved issue. It is proposed that the pairing is
mediated by the antiferromagnetic (AF) spin fluctuations resulting
in a S$\pm$ pairing \cite{MazinPRL,KurokiPairing}, which now
becomes a widely perceived picture. The experimental evidence that
directly supports S$\pm$ pairing is still quite rare. One was
drawn from the scanning tunnelling spectroscopy (STS) and the
quasi-particle interference measurements in
Fe(Se,Te)\cite{HanaguriScience}. Another indirect evidence is the
observation of the resonance peak of the imaginary part of the
spin susceptibility at $(\pi, \pi)$ in the inelastic neutron
scattering experiment\cite{NeutronResonance}. Theoretically it was
argued that this type of pairing should also be fragile to the
non-magnetic impurities
\cite{KontaniSpm,OgataTheo1,BangImpuritySpm}. One of the puzzles
concerning this model is why superconductivity survives up to a
relatively high temperature in some doped samples, such as the
Ba(Fe$_{1-x}$T$_x$)$_2$As$_2$ (T=Co, Ni, Ru, etc.), where the
dopants are supposed to be the pair breakers. The spatial and
energy dependent LDOS induced by the impurity scattering can be
regarded as the {\it fingerprints} for a formal check on this
peculiar pairing model
\cite{KontaniSpm,OgataTheo1,BangImpuritySpm,KemperCodoped,NakamuraTheoImpurity,HuTheoImpurity}
which can be measured by the STS. After cleaving an iron-based
superconductor, one usually encounters a polar and messy
surface \cite{HasanARPES,ShanLNatPhys}, while an atomically resolved
image was sometimes observed in a reconstructed surface
\cite{MasseeSTM,HoffmanSTMReview,DavisSTMScience}. However, it is
very difficult to locate a {\it well-defined quantum impurity} and
investigate the related LDOS on such surfaces with an exception in
Ca(Fe$_{1-x}$Co$_x$)$_2$As$_2$\cite{DavisSTMScience}. In this
paper, we report the success in locating the Co impurities and
measuring the spatially resolved STS at different energies and
temperatures obtained from the Na(Fe$_{1-x}$Co$_x$)As single
crystals. Our results indicate an unexpected weak variation of the
STS and LDOS when going through a Co impurity site. Combined with
the theoretical calculations, we propose explanations for this
dichotomy (finite LDOS but weak variation). Our findings put
constraints on the theoretical models of the pairing state.

\section{Experiments}

High quality Na(Fe$_{1-x}$Co$_x$)As single crystals were
synthesized by the flux method \cite{SampleGrowth}. The samples were cleaved at room
temperature in an ultra-high vacuum with a base pressure better
than $2\times10^{-10}\;$Torr, then transferred into the scanning
tunneling microscopy (STM) head immediately and cooled to a
desired temperature. The STS were measured with an ultra-high
vacuum, low temperature and high magnetic field scanning probe
microscope USM-1300 (Unisoku Co., Ltd.). In all STM/STS
measurements, Pt/Ir tips were used. The surface topographies were
recorded using a bias voltage of $V_\mathrm{bias} = 40\;$mV and
tunneling current of $I_\mathrm{t} = 100\;$pA. To reduce the noise of
the differential conductance spectra, a lock-in technique with an
ac modulation of $0.1\;$mV at $987.5\;$Hz was typically used.

\section{Results}

\subsection{Sample characterization}

\begin{figure}
\includegraphics[width=8cm]{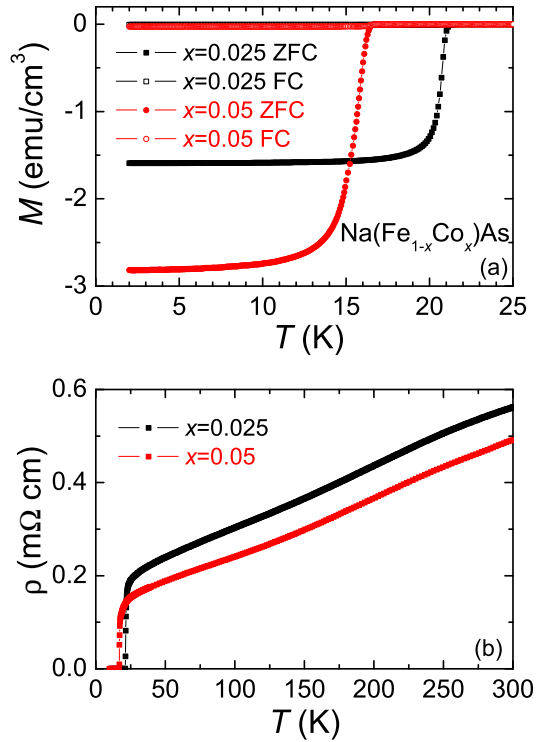}
\caption{(Color online) (a) Temperature dependence of volume
magnetization of Na(Fe$_{0.975}$Co$_{0.025}$)As and Na(Fe$_{0.95}$Co$_{0.05}$)As single crystals after zero-field cooling
(ZFC) and field cooling (FC) at 10$\;$Oe. The difference in the
ZFC magnetization of the two samples comes from the different
demagnetizing effect. (b) Temperature dependence of resistivity for the two Na(Fe$_{1-x}$Co$_x$)As samples with different doping levels.} \label{fig1}
\end{figure}

Figure~\ref{fig1}(a) shows the temperature dependence of the volume dc
magnetization ($M$) after zero-field-cooling (ZFC) and field cooling (FC) processes.
The magnetization on the samples of two doping levels showed sharp
superconducting transitions with $T_\mathrm{c} \approx 16.8\;$K for $x = 0.05$ and
$T_\mathrm{c} \approx 21.2\;$K for $x = 0.025$. Since the single crystals are thin and plate like, the
demagnetizing factor is relatively large; that is the reason the
ZFC magnetization values of these two samples are different.
The error in the measurement of dimensions, especially the
thickness, could give some error in the calculation of the ZFC
susceptibility. Figure~\ref{fig1}(b) shows the resistive transitions of the two samples. It is clear that the sample with more charge doping level (x=0.05) is lower than that at the optimal doping (x=0.025) although their T$_c$ values are opposite. The different values of the resistivity at 300$\;$K may be related to the charge doping effect in the samples.

\subsection{Identification of Co dopant atoms in Na(Fe$_{1-x}$Co$_x$)As}

\begin{figure}
\includegraphics[width=9cm]{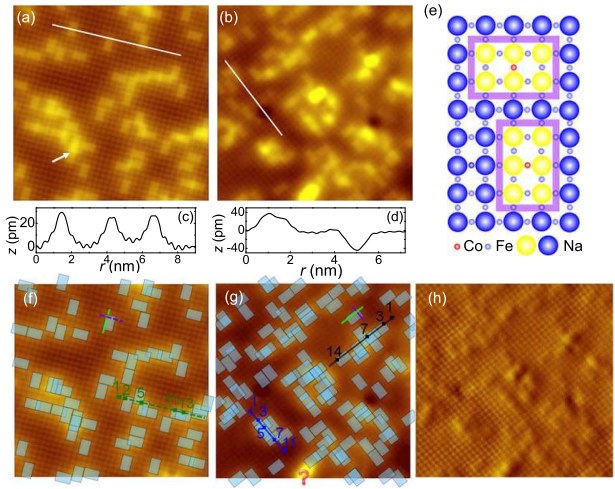}
\caption{(Color online) (a, b) High-resolution topography in a region of
14$\times$ 14 nm$^2$ on Na(Fe$_{1-x}$Co$_x$)As samples with
$x=0.025$ (a) and $x=0.05$ (b). (c, d) The spatial dependence of
height $z$ measured along the white lines marked in (a) and (b).
(e) Illustration for atoms near the Co-impurity sites. It is clear
that each Co atom neighbors with six Na atoms (highlighted by the
yellow circles), forming the $2\times1$ rectangular block (see
text). (f, g) Illustration for counting the $2\times1$ blocks on (a, b),
and the $2\times1$ block patterns are marked with light-blue
rectangles. (h) The differential picture of (b) which is helpful to make the counting.
} \label{fig2}
\end{figure}

Figures~\ref{fig2}(a) and \ref{fig2}(b) present the topographic image of
the cleaved surface of samples with (a) $x=0.025$ and (b)
$x=0.05$. The background shows a well-ordered square lattice with
a constant of $3.80\; {\AA}$. Because the Na atoms are
distributed equally into two layers between two neighboring
As-layers, the cleaving occurs between the two Na layers leading
to an un-polar (charge neutral) surface \cite{HessSTMPRL}. In both
samples, beside the observed square lattice, one can see some
rectangular blocks with two unit cells constructed by six Na atoms
(hereafter named a $2\times1$ block) which are a bit brighter
than the background as shown in Figs.~\ref{fig2}(a) and 2(b).
These blocks align along either [100] or [010] directions. The
brightness is enhanced when many blocks overlap each other [marked
by an arrow in Fig.~\ref{fig2}(a)]. After considering the
arrangement of Na atoms in one $2\times1$ block, we argue that
each block is corresponding to one individual Co impurity atom.
Two reasons can be given to support this argument, as addressed below.

\begin{figure}
\includegraphics[width=8cm]{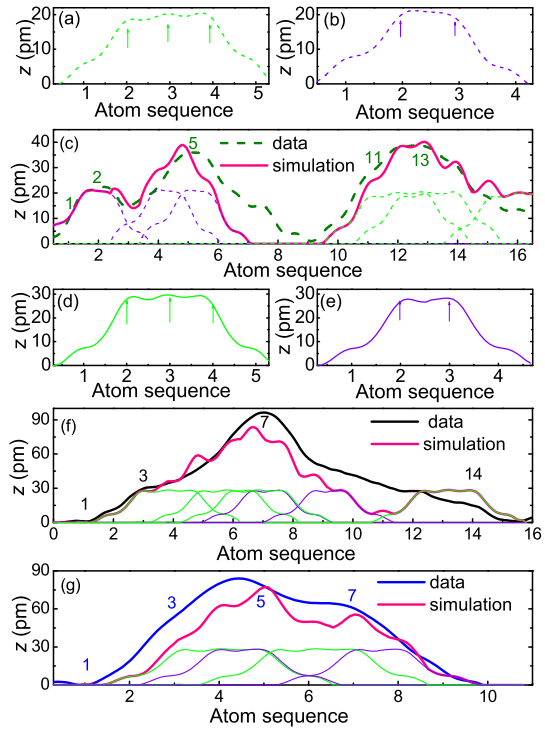}
\caption{(Color online) Simulations of the height on overlapping of $2\times1$
blocks. (a),(b),(d) and (e) are line profiles of a single block
marked by green and violet dashed/solid lines in Figs.~\ref{fig2}(f) and (g), while
(c), (f) and (g) show three simulated line profiles of combined altitude
by overlapping blocks illustrating in Figs.~\ref{fig2}(f) and (g).
One can find that the simulated lines match well with the experimental data. } \label{fig3}
\end{figure}

Firstly, when one Co atom is doped to the Fe site, it naturally
neighbors with six Na atoms on the surface layer, forming a
2-unit-cell rectangular ($2\times1$) block as illustrated by
Fig.~\ref{fig2}(e), the orientation of the $2\times1$ block is
determined by the selective positioning of the Co atom on the
$a$-axis or $b$-axis of the Fe-As square lattice. Secondly,
counting the ratio between the number of the typical $2\times1$
blocks and the total number of Fe and Co atoms, we find that it is
very close to the doped concentration of Co, i.e., the nominal
value of Co/(Fe+Co) in the samples. We counted the number of
$2\times1$ blocks as shown in Fig.~\ref{fig2}(f) and (g). The very
bright regions on the over-doped Na(Fe$_{0.95}$Co$_{0.05}$)As
sample are originated from overlapping of several $2\times1$
blocks which can be distinguished from the differential figure as
shown in Fig.~\ref{fig2}(h). The stacking blocks are distinguished
from different height of the Na atoms and contrast level of the
experimental image. The quantities of Co atoms counted on
Na(Fe$_{0.975}$Co$_{0.025}$)As and Na(Fe$_{0.95}$Co$_{0.05}$)As
samples are 64 and 110 respectively with the counting error of
10\%, while the total number of Fe sites should be 2677 in such
14$\;$nm$\times$14$\;$nm area. Only one bright region with the
question mark at the bottom edge on the
Na(Fe$_{0.95}$Co$_{0.05}$)As sample as shown in Fig.~\ref{fig2}(g)
hasn't been taken into account, due to the uncertainty. The
corresponding ratio of doped Co atoms over all the Fe+Co sites
Co/(Fe+Co) on the sample with $x=0.025$ $\approx 2.39\pm0.24\,\%$;
while on the sample $x=0.05$, we get Co/(Fe+Co) $\approx
4.11\pm0.41\,\%$. The two calculated ratios are comparable with
their chemical doping levels 2.5\,\% and 5\,\%, respectively. This
could not be achieved by accident. We also checked the height
where several blocks overlap each other. As shown in
Figs.~\ref{fig3} (a), (b), (d), and (e), we present the height
distributions crossing one single $2\times1$ block in two
perpendicular directions. We used these distributions as basic
cells to simulate the height where several blocks overlap by a
simple summation. The simulation is undertaken based on the
distribution of the $2\times1$ blocks shown in Fig.~\ref{fig2}(f)
and (g). The accumulated height equals to the summation of the
total contributions given by several blocks demonstrated by the
frames. We test this simple method along three typical lines in
Fig.~\ref{fig2}(f) and (g). One can see that the consistency
between the simulated results and the data is remarkable. This
clearly convinces that the very bright spots are constructed by
several $2\times1$ blocks overlapping each other, which validates
the counting method we proposed here. This result may also suggest
that the local work function on six Na atoms of the $2\times1$
blocks is reduced, which results in a linear enhancement of the
height when the blocks overlap each other. Such conclusion can be
derived by the expression of tunneling current if the second order
small quantities in the Taylor expansion are omitted.

Recently, Erwin and Mazin have performed first principles
calculations of the map of tunnelling current in NaFeAs with an
isolated surface Co impurity \cite{MazinDiscussion}, and found
that the calculated topographic image closely resembles ours and
reproduces the $2\times1$ block that we assigned to a Co impurity.
These $2\times1$ blocks seem to be slightly higher (about 30
pm) than the surrounding atoms, as shown by the line-cut in
Figs.~\ref{fig2}(a) and (b) with the landscape shown in
Figs.~\ref{fig2}(c) and (d), respectively. At some positions, we
see dark areas characterized by a lower height, as shown by the
line-cut in Figs.~\ref{fig2}(b) and (d). These are probably
induced by the vacancies of the atoms beneath the surface layer.

\subsection{Impurity state on Co-dopant site}

\subsubsection{STS far away from the Co-impurity}

\begin{figure}
\includegraphics[width=8cm]{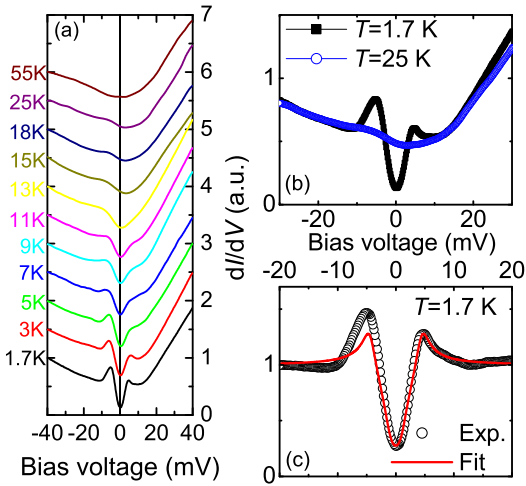}
\caption{(Color online) (a) The STS measured on
Na(Fe$_{0.95}$Co$_{0.05}$)As at a position far away from the
Co-impurities at temperatures from 1.7 K to 55 K. (b) STS taken at 1.7 K
and 25 K. (c) The STS measured at 1.7 K
normalized by that of 25 K (symbols). The solid line is a fit
according to the Dynes formula with a gap of 4.5 meV and a
scattering rate of $\Gamma= 0.7\;$meV.} \label{fig4}
\end{figure}

\begin{figure}
\includegraphics[width=8cm]{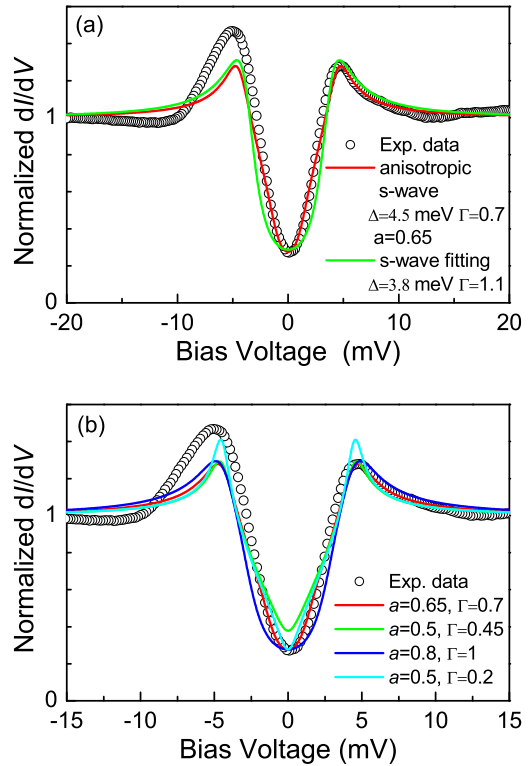}
\caption{(Color online) STS measured on Na(Fe$_{0.95}$Co$_{0.05}$)As
at 1.7 K and the fitting with different gap functions (a)
and different values of fitting parameters for anisotropic s-wave gaps (b). } \label{fig5}
\end{figure}

In the sample with $x = 0.05$ we measured the evolution of
tunnelling spectrum with temperature and showed it in
Fig.~\ref{fig4}(a). At low temperatures, two coherent peaks near
the gap edges can be clearly seen. The asymmetric structure of the
STS curve is induced by the background, which can be evidenced by
the data measured at 25 K ($> T_\mathrm{c} = 16.8\;$K) as shown in
Fig.~\ref{fig3}(b). In Fig.~\ref{fig4}(c), we present the
experimental STS at 1.7 K that normalized by the data at 25 K. On
the STS curve, there is a finite value of zero-bias conductance,
which seems to be a common feature of Co-doped samples
\cite{HoffmanPRL,MasseeSTMPRB,YehSTMPRL}, and will be addressed
later. No feature of two gaps is observed here. This is consistent
with the angle-resolved photo-emission spectroscopy measurements
on the same sample, which indicates the same gaps, $\sim5\;$meV,
on both the electron and hole pockets.\cite{Co111ARPES} The
normalized data was fitted with the Dynes model\cite{DynesFit}
with the gap function $\Delta(\theta)=\Delta[(1-a)\cos2\theta+a]$
to fit the experimental data. First we try the isotropic S-wave
with anisotropic factor $a=1$, as shown in the Fig.~\ref{fig5}((a). The result is
shown by the green curve. In Fig.~\ref{fig5}(b), we show the
fitting curves with different anisotropy $a$ and $\Gamma$. It is
found that an optimized fitting can be obtained when using $a=0.65$,
i.e., $\Delta(\theta)=4.5(0.35\cos2\theta+0.65)\;$meV and scattering
rate $\Gamma=0.7\;$meV. One point we should mention is that,
there is a slight enhancement of $\mathrm{d}I/\mathrm{d}V$ at about -5 meV even
above $T_\mathrm{c}$, e.g., at 25 K. This feature is smeared out
at elevated temperatures and vanishes at about 55 K. Interestingly
this feature seems to be linked with the asymmetry of the STS, for
example, the coherence peak at $-\Delta = -4.5\;$meV is higher
than that at $+\Delta = 4.5 \;$meV, even at 1.7 K. The
relationship between the high temperature feature and the
asymmetric superconducting coherence peaks remains to be
understood.

\subsubsection{STS across a Co-impurity}

\begin{figure}
\includegraphics[width=9cm]{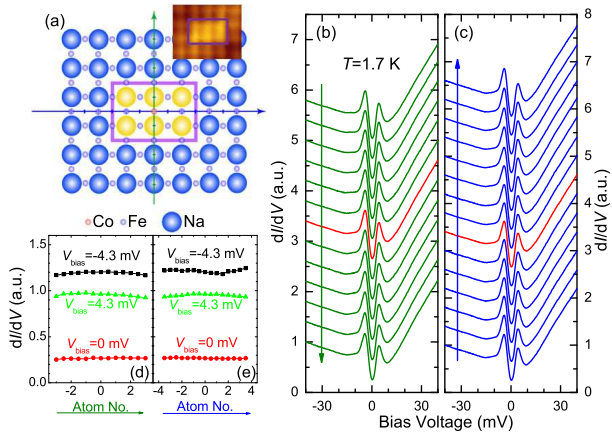}
\caption{(Color online) (a) Illustration of the atomic structure
based on the topographic image (shown by the inset) of
Na(Fe$_{0.95}$Co$_{0.05}$)As at 1.7 K, with a Co-impurity
highlighted by a red circle surrounded by Na atoms (big yellow and
blue circles). (b, c) Spatially resolved STS measured along the
two arrowed lines in (a) with a spacing of every half lattice
constant. The red lines show data measured just at the center
of the $2\times1$ rectangular block. (d, e) The value of $\mathrm{d}I/\mathrm{d}V$
determined near the gap edge (-4.3 meV and 4.3 meV) and zero bias
when crossing the Co-impurity point.} \label{fig6}
\end{figure}

In order to have a detailed comparison between the LDOS patterns
as expected by the theoretical models and that induced by the
Co-impurity, we select a $2\times1$ block and do the spatial
resolved STS measurements with spacing of every half lattice
constant, as marked by the dots in Fig.~\ref{fig6}(a). The raw
data of STS are shown in Figs.~\ref{fig6}(b) and \ref{fig6}(c) for the two
perpendicular directions. The data measured right on top of the
Co-impurity is shown here by a red curve. It is astonishing that
the shape of the STS does not change in any noticeable way upon
moving through the Co-impurity. The value of $\mathrm{d}I/\mathrm{d}V$
taken from these STS curves at three typical voltages
($V_\mathrm{bias}= -4.3$, 0, and 4.3 mV) are shown in
Figs.~\ref{fig6}(d) and \ref{fig6}(e) with respect to the positions. One can
clearly see that the LDOS at either zero energy or the gap edges
are quite uniform.

\begin{figure}
\includegraphics[width=9cm]{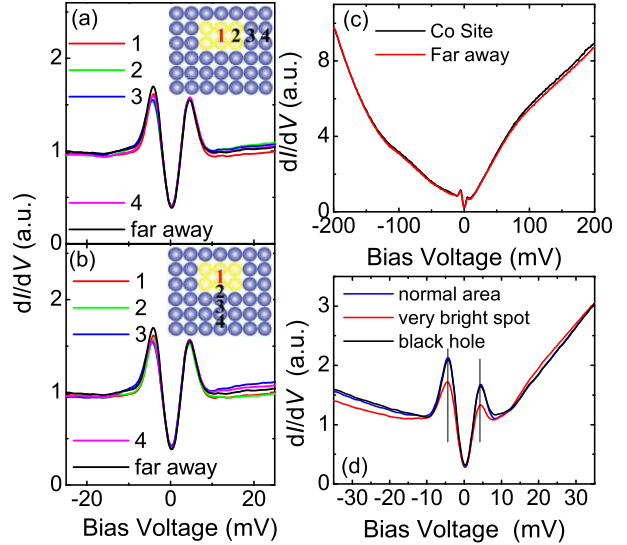}
\caption{(Color online) (a, b) The tunnelling spectra
normalized by the normal state one measured at different
positions: 1, on site of the impurity, 2-4 corresponds to those
measured at the 2nd, 3rd and 4th lattice constants away in two
perpendicular directions. (c) STS measured on Na(Fe$_{0.95}$Co$_{0.05}$)As
at 1.7 K. It is found that the STS measured just on the Co site is very
similar to the one measured far away and it has no special feature in
the voltage range from -200 mV to 200 mV besides the superconducting
gap at low energies. An asymmetric background is seen here which may
be induced by the multi-band and the band edge effect. (d) STS measured
on a very bright spot, a black hole and somewhere far away from such features.
One can see that the energy gap values are almost the same at these specific locations.} \label{fig7}
\end{figure}

In order to examine the spatial dependence of the spectrum on the
Co site and nearby, we show the spectra at different positions in
Fig.~\ref{fig7}(a) and \ref{fig7}(b). We also measured
the STS just on the Co site in a large bias voltage
range as shown in Fig.~\ref{fig7}(c). One can find an asymmetric
background which is similar to the spectra measured on LiFeAs\cite{HessSTMPRL}.
The theoretical simulation indicates that there are clear Co dopant
induced resonances at -800 mV and 200 mV in BaFe$_2$As$_2$\cite{KemperCodoped},
and some special characteristics at about 150 mV was observed and
suggested to image Co atoms on the surface of slightly Co-doped
CaFe$_2$As$_2$\cite{DavisSTMScience}. However we didn't find
any characteristic features within $\pm200\;$meV at either the
Co site, or somewhere far away from the Co sites. To investigate
the spectrum affected by the black holes and the very bright spots
originated from the overlapping of several $2\times 1$ blocks in Fig.~\ref{fig2}(b),
we measured the spectra on such places and showed them in Fig. ~\ref{fig7}(d).
One can see that the gap values as well as the bottom heights
are almost the same at different positions. The superconducting
coherence peaks on the spectrum measured on a bright spot are
suppressed comparing to the one measured at the point without any features. The black holes seem to affect the spectrum very
little which suggests that it arises from the depletion of some Na
atoms on the top surface.

\subsubsection{Theoretical simulated results of the LDOS on the Co site}

In order to understand the spatial dependence of the spectrum
on the Co site and nearby in our experiments, we did theoretically
calculations based on the S$\pm$ and S++ models. First we analyze
the gap functions of the S$\pm$ wave superconductivity obtained
in random phase approximation (RPA). We use the multi-orbital
Hubbard model proposed by Kuroki et al.\cite{KurokiPairing},
which is down-folded from the first-principle calculation. In
this scheme, five Fe 3d orbitals of $d_{3z^2-r^2}$, $d_{zx}$,
$d_{yz}$, $d_{x^2-y^2}$, and $d_{xy}$ are kept, and the hopping
integrals between these orbitals for the LaFeAsO (1111) system is used.
This will be justified since the Fermi surface for NaFeAs near
$k_z=0$ is similar to that of LaFeAsO, although NaFeAs has stronger
three-dimensionality than 1111 system. The effect of the $z$-direction
hopping on the impurity scattering is a future problem. In this paper,
we do not go into details of the material-dependent model, but instead we compare
the experiment and theory in a simple model which contains essence
of the multi-band unconventional (S$\pm$ and S++) superconductivity.
It is found that, when we use orbital representation instead of
band representation, the gap functions obtained in RPA can be
reproduced quite well by using the real-space short-range pairings
(up to next-nearest-neighbor sites). The details are explained elsewhere.\cite{OgataTheo2}
Note that, in the RPA calculation around $T_\mathrm{c}$, the absolute
values of the gap functions are not determined, and only their
relative sizes are obtained. Therefore, we adjust the largest gap value
to be consistent with the experiment. On the other hand, for the S++ case,
the gap functions are obtained from the analysis of the five-orbital
attractive Hubbard model.

The impurity is treated as a single site in the two-dimensional square
lattice with an on-site impurity potential. The orbital dependence of
the potential is neglected. The value of the impurity potential is chosen as
$I=1.0\;$eV \cite{KemperCodoped} and $I=-0.35\;$eV \cite{NakamuraTheoImpurity}
according to the previous first-principle calculations. The gap functions
around the impurity site are assumed to be spatially uniform for simplicity.
The effect of this simplification is discussed shortly. The actual calculation
of LDOS is performed using Chebyshev polynomial expansion method \cite{ChebyshevTheory}
with $900\times900$ lattice sites and 3600 polynomials. The Lorentzian-type broadening
kernel is used and the broadening factor is determined so as to make $\gamma/\Delta_\mathrm{max}$
consistent with experiment. Because of this broadening, LDOS at $E=0$ remains finite
even in the S++ case, which comes mainly from the tail of the peak at the gap edge.
The electron number is fixed at $n=6.05$ by adjusting the chemical potential.

There are two important factors that should be included when we compare the
theoretical calculations with experimental data. First, there is a band
renormalization effect due to electron-electron correlation which is
neglected in the first-principle calculations. In iron pnictides,
the effective mass obtained in ARPES is larger than that obtained in the
first-principle calculation by a factor of about 2.5. Therefore, in order
to take account of the band renormalization factor (not included in the
first-principle calculations), the hopping integrals and site-potentials
in the model Hamiltonian should be reduced by a factor 2.5. By considering
this effect, we reduce the energy scale of the obtained results by factor
2.5, for simplicity. The other important factor is that the experimental tunneling spectra were
taken under the condition of constant tunneling current. In all the experiment
measurements, the spectra were taken with the same initial tunneling current
(e.g., $100\;$pA) at some bias voltage $V_0$ (e.g., $40\;$mV). So there was
a normalization process using the integral area of sample LDOS from 0 to $V_0$
of each curve if we simply considered the tip LDOS as a constant.\cite{HoffmanSTMReview}
Such normalization makes very little change of the tip height which can be
omitted because the tunneling current decays exponentially with increasing
of height. This means that LDOS obtained theoretically should be normalized
in accordance with the experimental condition of constant current. This is
carried out by rescaling the amplitude of LDOS for each site so that the
energy integral of LDOS between $E=0$ to -20 meV becomes a constant for
the negative parts of calculations have fewer features.

\begin{figure}
\includegraphics[width=8cm]{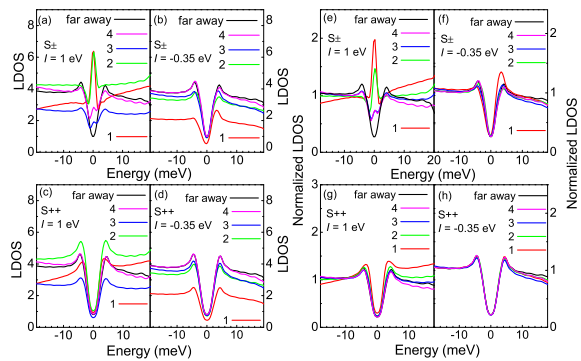}
\caption{(Color online)  (a-d) Un-normalized STS calculated in a simplified
model based on S++ and S$\pm$ pairing gaps at several sites near the impurity
with the same spatial registration as the experiment (based on the band
parameters of the LaFeAsO system \cite{OgataTheo2}) (e-f)
Curves of (a-d) normalized by the integral area of LDOS from -20$\;$mV
to 0$\;$mV in corresponding to the constant-current situation
in the experiment. } \label{fig8}
\end{figure}

After these treatments, LDOS shown in Figs.~\ref{fig8}(e)-\ref{fig8}(g) are obtained.
For the case of S$\pm$ and scatting potential $I = 1.0\;$eV [Fig.~\ref{fig8}(e)]
and at the impurity site, the simulated data indicate a strong in-gap bound
state peak at zero energy as shown in the previous paper\cite{OgataTheo1}.
For other cases, the bound state peaks are absent and shape of the spectrum
changes when it is moving away from the impurity site. From Figs.~\ref{fig8}(f)
and \ref{fig8}(h), we can see that the results with $I=-0.35$ eV for
S++ (Fig.\ref{fig8}(h)) are quite consistent with the experimental data.
However, we cannot exclude the possibilities of S$\pm$ model with
$I=-0.35$ eV and the S++ model with $I=1$ eV, since the simulated
results and experimental data are not very distinct.

\begin{figure}
\includegraphics[width=8cm]{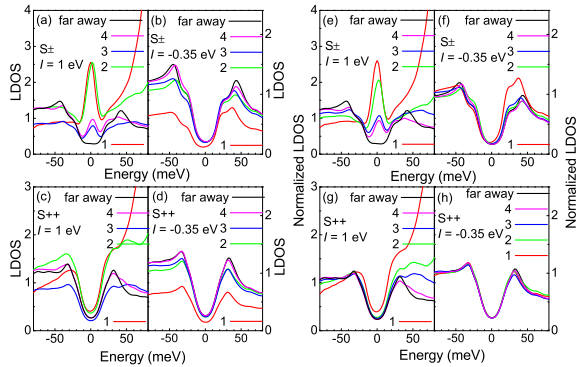}
\caption{(Color online) (a-d) The STS calculated in a five-orbital
model based on S++ and S$\pm$ pairing gaps at several sites near
the impurity with the same spatial registration as the experiment
(based on the band parameters of the LaFeAsO system \cite{OgataTheo2}) (e-f)
Curves of (a-d) normalized by the integral area of LDOS from -80$\;$mV
to 0$\;$mV in corresponding to the constant-current situation
in the experiment.} \label{fig9}
\end{figure}

At the end of this part, we want to check the validity of neglecting the spatial
dependence of the gap functions around the impurity site. For this purpose,
we have to determine the gap functions around the impurity self-consistently
by solving the Bogoliubov-de Gennes equation iteratively just as in Ref.~\cite{OgataTheo1}.
However, a large numerical cost of solving Bogoliubov-de Gennes equation limits
the accessible energy scale. If we try to reproduce the experimental gap size,
the self-consistency calculation becomes unstable or unrealistic suffered from
the finite-size effect of the available lattice. This is just a technical reason,
but we have to assume large attractive interactions in order to obtain self-consistent
solutions. In this calculation, we solve Bogoliubov-de Gennes equation used in Ref.~\cite{OgataTheo1}
that is constructed so as to reproduce the RPA S$\pm$ results and contains
attractive interactions up to the next-nearest-neighbor sites. On the
other hand, for the S++ case, we use a model with short-range attractive
interaction. We use a lattice with $28\times28$ sites with an impurity at
the centerin order to obtain eigenvalues and eigenfunctions. For the
LDOS calculation, we use the ``supercell'' method in which $28\times28$-site
lattice with an impurity is treated as a ``unit cell'' and the whole system is
assumed to be composed of a $13\times13$ repeat of this unit cell.\cite{OgataTheo1}
Figures~\ref{fig9}(a)-\ref{fig9}(d) show the LDOS obtained by this
self-consistent calculation. Since the attractive interaction is large,
the obtained gap function is larger than the experimentally observed gap values.
Note that LDOS is not normalized and the band renormalization factor is not
taken into account. For comparison, we show the LDOS without self-consistent
calculation in Figs~\ref{fig8}(a)-\ref{fig8}(d), which are identical with
Figs.~\ref{fig8}(e)-\ref{fig8}(h) but they are not normalized. From this
comparison, we can see that the results do not change qualitatively even if
we assume a site-independent gap functions [Figs~\ref{fig8}(a)-\ref{fig8}(d)]
compared with the full self-consistent calculation in Figs.\ref{fig9}(a)-\ref{fig9}(d).
One dominant difference is a rather large peak at $E>50\;$meV in
Figs.\ref{fig9}(a) and \ref{fig9}(c) (i.e., for $I=1.0\;$eV case) on the
impurity site (line 1: red curve). We find that this large peak is due to
the impurity-induced resonant state that exists even in the normal state.
Although the impurity-induced resonant states exist in Figs.~\ref{fig8}(a)
and \ref{fig8}(c), it is located outside of the energy-range shown here.

\subsection{Discussion on the Impurity effect by Co doping}

\begin{figure}
\includegraphics[width=9cm]{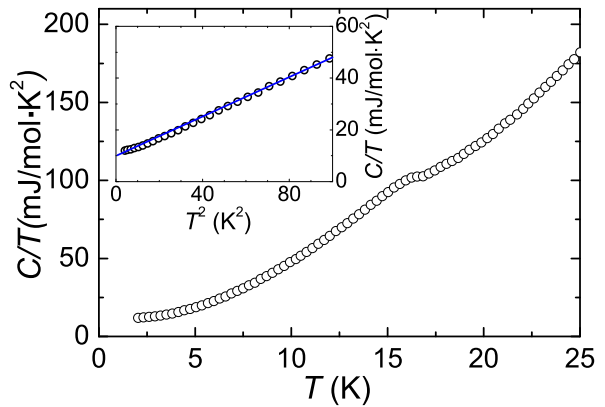}
\caption{(Color online) Temperature dependence of specific heat of
Na(Fe$_{0.95}$Co$_{0.05}$)As. The inset shows an enlarged view of
the same data in the low temperature limit.} \label{fig10}
\end{figure}

The experimentally measured STS seems to change very weakly when
crossing the Co site as mentioned above. On one hand, we don't see
the sharp bound state peaks at the impurity site as theoretically
predicted for the S$\pm$ pairing gap for $I = 1$ eV with the
non-magnetic impurities. On the other hand, we do see a continuum
of in-gap states, which lifts the zero-bias conductance to a
finite value. Theoretically we find that the effect of Co-impurity
on LDOS is similar both for S$\pm$ and S++ case when $I =
-0.35\;$eV. One may argue that the scattering effect given by the
Co-impurity is very weak i.e., $I \sim 0$. However, specific heat
data shows a significant residual specific heat coefficient
$\gamma_0=9.8$ mJ/mol$\cdot$K$^2$ (as shown in Fig.~\ref{fig10})
suggest that the scattering in the Co-doped case is significant.
This phenomena occurs in other doped samples as well when the
dopants go directly to the Fe sites \cite{FangPRBCo122}. The
density functional calculation of the local substitution of Co or
Ni for Fe suggest that no extra charge carriers are doped into the
system\cite{SawatzkyTheo}. In this case, the Co-impurity should
behave as an impurity scatterer for Cooper pairs \cite{LevyRPES}.
Provided the expectable influence given by the Co-impurities, it
is paramount that we understand the lack of the spatial variation
of the LDOS spectrum nearby the Co-site within the S$\pm$ model.

Now we discuss the possible reasons for this dichotomy, and
constraints given by our results on the theoretical pairing
models. Although there is a discrepancy (especially near the gap
edge) between the experimental data and the theoretical results,
the S++ model would give a much better consistency. However, the
S$\pm$ model may still be relevant in the following circumstances:
(1) The scattering potential $I$ is around -0.35 eV, as shown in
Fig.~\ref{fig8}(f). In the absence of precise independent
assessment on the scattering potential, we assume that the
scattering potential is relatively weak. (2) The Co-impurity may
behave as an extended scattering center, so that the scattering
between the electron and hole Fermi pockets (if would be the cause
of pairing) will not be affected due to the finite range. Since
the inter pocket scattering is the major cause for the pairing
based on the S$\pm$ model, that may be the reason why the
superconductivity survives in the presence of such impurities.
With this picture, the obvious scattering as revealed by the
residual resistivity and specific heat data are probably induced
by the intra-pocket scattering which can suppress the
superconductivity moderately, but not strongly. Such scattering
results in the expectation of in-gap filled states with no sharp
resonance states seen. (3) The last possibility would be that
the pairing is induced by strong local super-exchange effect\cite{SiQMPRL,HuTwoOrbital}
which will certainly not be weakened so much by the non-magnetic
impurity and relatively weak scattering.

\section{Conclusions}
In summary, we successfully locate the doped Co atoms in
Na(Fe$_{1-x}$Co$_x$)As using scanning tunnelling microscopy.
Negligible spatial variation of the tunnelling spectrum near the
Co impurity is observed, which is in sharp contrast with the
theoretical simulations with the S$\pm$ model taking $I=1$ eV. We
explain our data with the S$++$ or S$\pm$ models. The S$++$ model can be consistent with the data, however, the S$\pm$ model is still relevant in the following three cases: (1) The scattering potential I given by a Co dopant is relatively small,  for example, \emph{I} ~ -0.35 eV; (2) The Co-dopants give rise to the extended scattering centers which will not significantly affect the interpocket scattering; (3) The pairing may be induced by local strong magnetic super-exchange effect.

\section*{Acknowledgments}
This work is supported by the Ministry of Science and Technology
of China (973 projects: 2011CBA00102, 2012CB821403,
2010CB923002), NSF and PAPD of China. Work at Los Alamos was supported by the
U.S. DOE under contract No. DE-AC52-06NA25396 and the Office of
Science (BES).

$^*$ hhwen@nju.edu.cn

\end{document}